\documentclass[twocolumn,amsmath,amssymb,aps,pre]{revtex4-1}

\usepackage[pdftex]{graphicx}
\usepackage{amssymb,amsfonts,amsmath}
\usepackage{color}
\usepackage[usenames,dvipsnames]{xcolor}

\definecolor{darkchocolate}{rgb}{0.55, 0.27, 0.07}

\begin{document}


\title{Equilibrium and dynamic pleating of a crystalline bonded network}

\author{Saswati Ganguly}
\author{J\"urgen Horbach}
\affiliation{Institut f\"ur Theoretische Physik II: Weiche Materie, Heinrich 
Heine-Universit\"at D\"usseldorf, Universit\"atsstra{\ss}e 1, 40225 D\"usseldorf, 
Germany}
 
\author{Peter Sollich}
\affiliation{King's College London, Department of Mathematics, Strand, London WC2R 2LS, U.K.}
\author{Parswa Nath}
\affiliation{TIFR Centre for Interdisciplinary Sciences, 21 Brundavan Colony, Narsingi, Hyderabad 500075, India} 
\author{Smarajit Karmakar}
\affiliation{TIFR Centre for Interdisciplinary Sciences, 21 Brundavan Colony, Narsingi, Hyderabad 500075, India} 
\author{Surajit Sengupta}
\affiliation{TIFR Centre for Interdisciplinary Sciences, 21 Brundavan Colony, Narsingi, Hyderabad 500075, India}


\date{\today}


%
\begin{abstract}
We describe a phase transition that gives rise to structurally
non-trivial states in a two-dimensional ordered network of particles
connected by harmonic bonds. Monte Carlo simulations reveal that
the network supports, apart from the homogeneous phase, a number
of heterogeneous ``pleated'' phases, which can be stabilised by an
external field. This field is conjugate to a global collective variable
quantifying ``non-affineness'', i.e.~the deviation of local particle
displacements from local affine deformation. In the pleated phase,
stress is localised in ordered rows of pleats and eliminated from the
rest of the lattice. The {\em kinetics} of the phase transition is unobservably slow in molecular dynamics simulation near coexistence, due to 
very large free energy barriers. When the external field is increased further to lower these barriers, 
the network exhibits rich dynamic behaviour: it transforms into a {\em metastable}
phase with the stress now localised in a {\em disordered} arrangement of
pleats. The pattern of pleats shows ageing dynamics and slow relaxation
to equilibrium. Our predictions may be checked by experiments on tethered
colloidal solids in dynamic laser traps.
\end{abstract}

\maketitle

\section{Introduction}
Fabricating  complex shapes by folding or pleating a two-dimensional
elastic manifold has recently emerged as a viable technological
paradigm, applicable over a large range of length scales, from microns
to nanometers~\cite{origami1,origami2,origami3,irvine}. A number of
these innovative ideas are equally applicable for atomic crystals or
for larger assemblies involving functionalized colloidal particles
joined together using polymer tethers~\cite{carpets} or micron-sized
lipid droplets~\cite{bayley,shape}. To make such attempts feasible, one
needs efficient ways to control local structural properties, preferably
in a reversible way. Thus, microscopic understanding of the underlying
thermodynamics and kinetics of these or similar local shape changes
would be valuable.

In this paper, we study in detail equilibrium and dynamic aspects of
a transition in a tethered network of colloidal particles -- from a
homogeneous to a heterogeneous phase containing an ordered arrangement
of {\em pleats}.  Within these pleats, the network penetrates itself,
with parallel rows of vertices folding back to completely or nearly
overlapping adjacent rows. These complex structures arise here {\em
spontaneously} as a result of an underlying,  equilibrium ``first-order''
phase transformation~\cite{CL}.

We show that pleating excitations are induced and controlled by a novel
external field conjugate to a collective variable~\cite{falk} defined as
follows. In earlier work, it was demonstrated that local displacements of
particles in a crystal away from their ideal positions may be decomposed
into affine and non-affine components~\cite{sas1,sas2,sas3}. External
stress couples to the affine part of the displacements. In analogy, one
imagines an external field, which is conjugate to a global collective
coordinate $X$, measuring non-affineness, to be defined explicitly
later. This ``non-affine'' field, $h_X$, is realisable experimentally
for colloidal solids using dynamic laser traps~\cite{sas2,HOT}. For
small $h_X > 0$, non-affine displacements, and consequently $X$, are
enhanced in a controlled manner computable within a linear response
framework. In a non-bonded conventional crystal, this leads to the
creation of defects~\cite{sas2}. The present paper is devoted to an
analysis of the consequences of large positive $h_X$ in a connected
network much beyond the linear response regime.

We consider a periodic lattice of point vertices in two dimensions
(2d) that are connected to their nearest neighbours by harmonic
springs. Crystalline 2d networks have been studied extensively in the
past~\cite{network, pleats1,pleats2}, partly because of their biological
significance (e.g.~as a simple model for the spectrin network in red
blood corpuscles)~\cite{spectrin, spectrin1}. Note that such a network
has non-trivial properties, such as the presence of pleats, only when
the bond length is random or exceeds the nearest neighbour distance
set by the density~\cite{pleats2} by a critical amount. In this case
the network suffers an instability and catastrophic collapse. This is
in stark contrast to the equilibrium first order transition~\cite{CL}
in the presence of $h_X$ (and zero external stress) to an {\em ordered}
pleated state that we describe, we believe for the first time, in the
present work. For this transition, we obtain the relative free energies
of the pleated phases as well as individual free energy barriers in
quantitative detail -- a first but essential step towards efficient
control over their formation.

We use two distinct particle-based simulation techniques for this study.
Monte Carlo (MC) simulations~\cite{binder} in combination with sequential
umbrella sampling (SUS)~\cite{SUS} are used to compute the free energy
landscape as a function of $X$ at various values of $h_X$. We show
that SUS-MC simulations are able to detect metastable phases which are
inaccessible by conventional MC. Free energies of interfacial
structures between pleated and un-pleated regions of the lattice can also
be investigated and energy barriers for the formation of the product
phase determined. This advanced sampling method (SUS-MC) allows us
to not only identify this as a first order phase transition but also
to characterise the properties of the coexisting states separated by
interfaces. This latter point is remarkable because we obtain explicitly
an interface between an inhomogeneous state (i.e.~the pleated state) and
the normal crystal. Locating this interface is nontrivial and requires
the computation of local stresses.

We use molecular dynamics (MD) simulations~\cite{allen,frenkel} at
constant particle number, total area and temperature (NAT) to reveal
the kinetic aspects of this transition. In the MD simulation, the
transformation shows features that are different from the thermodynamic
first-order transition, as seen in SUS-MC. The transformation occurs at a
larger value of $h_X$ and leads to a {\em metastable} phase. We show, in
this work, the strong interplay between kinetics and thermodynamic phase
behaviour. The equilibrium pleated states are heterogeneous phases which are quite
difficult to realise via a kinetic pathway. Conversely, the nature of the metastable
states obtained in the kinetic transition can only be understood if one
is able to identify the underlying phase transition through SUS-MC.

We have carried out the simulations for both the pure, non-self avoiding
network as well as a model where the vertices of the network are
decorated with finite-sized colloidal particles. On a qualitative level,
the findings for both models are similar.

The rest of the paper is organised as follows. Section \ref{sec1}
introduces the local and global non-affine parameters, $\chi({\bf R})$ and $X$, as well as the
field $h_X$ conjugate to the latter. Moreover,
the details of the model solids and simulation methodology are
presented here. Then, in Sec.~\ref{sec2}, a comprehensive exposition of our
results for the network solid is given. We report results for analytic
calculations for small $h_X$, ground states and finite temperature phases
and the dynamical transition in this model. We end the paper with a
summary and conclusions as well as an outlook for future work.

\section{Models, Formalism and Simulation Details} 
\label{sec1}
In this section, we commence our discussion by first introducing
the non-affine field $h_X$ and the model Hamiltonians followed by a
description of the simulation methodologies used.

\subsection{The model Hamiltonian with the non-affine field}
Consider a reference configuration of $N$ particles where the particle
with index $i$ ($i=1,...,N$) is located at position ${\bf R}_i$.
A displacement of particle $i$ from its position on the reference lattice
is given by ${\bf u}_i = {\bf r}_{i}-{\bf R}_{i}$, with ${\bf r}_i$ the
instantaneous position of the particle. Now within a neighborhood $\Omega$
around particle $i$, we define relative atomic displacements ${\bf
\Delta}_{j} = {\bf u}_j-{\bf u}_i$ with particle $j\neq i \in\Omega$.
The ``best fit'' \cite{falk} local affine deformation ${\mathsf D}$ is the
one that minimizes  $\sum_j [{\bf \Delta}_{j} - {\mathsf D}({\bf R}_{j}
- {\bf R}_{i})]^2$ with the non-affinity parameter $\chi({\bf R}_i) >
0$ being the minimum value of this quantity. The minimisation procedure
amounts to projecting~\cite{sas1} ${\bf \Delta}_i$ onto a non-affine
subspace defined by a projection operator ${\mathsf P}$ such that
$\chi({\bf R}_i)= {\bf \Delta}^{\rm T}{\mathsf P}{\bf \Delta}$ where
$\Delta$ is the column vector constructed out of the  ${\bf \Delta}_i$. In
the projector ${\mathsf P} = {\mathsf I}-{\mathsf R}({\mathsf R}^{\rm
T}{\mathsf R})^{-1}{\mathsf R}^{\rm T}$, the $Nd\times d^{2}$ elements of
${\mathsf R}$ are given by ${\mathsf R}_{j\alpha,\gamma\gamma^{\prime}}
= \delta_{\alpha\gamma}R_{j\gamma^{\prime}}$ (here, the central particle
$i$ is taken to be at the origin). This projection formalism is perfectly
general and can be carried through for any lattice in any dimension.

Once the dynamical matrix ${\mathcal D}({\bf q}) = v_{BZ}^{-1}\langle
{\bf u}_{\bf q}{\bf u}_{\bf -q} \rangle$ is obtained ($v_{BZ}$ is the
volume of the Brillouin zone), we can calculate the ensemble average of
the non-affine parameter $\langle \chi \rangle$ using a coarse graining
procedure outlined in~\cite{sas1,sas2}. We include this in brief here
for completeness.

We define the coarse grained correlations, $C_{i\alpha,j\beta} = \langle
\Delta_{i\alpha}\Delta_{j\beta}\rangle$ where as before the Roman indices denote
particles and the Greek ones denote coordinates. The particles $i,j$ both
belong inside the neighborhood $\Omega$ of particle $0$.  Substituting
the definition of the displacement differences $\Delta_{i\alpha} =
u_{i}^{\alpha} - u_0^{\alpha}$ we obtain~\cite{sas1},
$$
C_{i\alpha,j\gamma} = \int \frac{d{\bf q}}{v_{BZ}}\,
{\mathcal D}^{-1}_{\alpha\gamma}({\bf q})(e^{i{\bf q}
\cdot{\bf R}_{j}}-e^{i{\bf q}\cdot{\bf R}_{0}})
(e^{-i{\bf q}\cdot{\bf R}_{i}}-e^{-i{\bf q}\cdot{\bf R}_{0}})
$$
The ensemble average of $\langle \chi \rangle$ is then given by $
{\rm Tr}~{\mathsf P}{\mathsf C }{\mathsf P}$, where ${\mathsf P}$ is
the projection operator. The non-trivial eigenvalues $\sigma_\mu$
of ${\mathsf P}{\mathsf C }{\mathsf P}$ and their corresponding
eigenvectors are the non-affine modes of the lattice. For example,
in the triangular lattice there are $8$ such modes when $\Omega$
corresponds to the nearest neighbour shell. The orthogonal subspace,
i.e.~the {\em affine} displacements, are spanned by the non-trivial eigenvectors
of $(1-{\mathsf P}){\mathsf C }(1 -{\mathsf P})$ and correspond to
the usual volumetric, uniaxial and shear strains together with local
rotations. The non-affine field does not affect the statistics of the
affine part of the displacements to linear order in $h_X$. Space-time
correlation functions of both affine and non-affine variables can also
be obtained using a similar procedure~\cite{sas1,sas2}.

In order to selectively excite lattice distortions that enhance
non-affine displacements, we introduce an extended microscopic
Hamiltonian~\cite{sas2} involving the thermodynamic conjugate
variables $h_{X}$ and $X$, with {\color{black} $X=N^{-1}\sum^N_i
\chi({\bf R}_i)$.} In analogy to conjugate variables like stress-strain
or pressure-volume, we add the product of $h_{X}$ and $X$ to
the Hamiltonian:
\begin{eqnarray} 
{\cal H} &=& {\cal H}_{0} - N h_{X}X \nonumber \\
&=& {\cal H}_{0} - h_X \sum_i^N\sum_{jk\in \Omega} 
({\bf u}_j-{\bf u}_i)^{\rm T}{\bf P}_{j-i,k-i}({\bf u}_k-{\bf u}_i).
\nonumber \\
\label{hamil_gen}
\end{eqnarray}
Here, ${\cal H}_{0}$ represents the Hamiltonian of any standard solid. 

The second term in Eq.~(\ref{hamil_gen}), $- N h_X X$, involves appropriate
Cartesian components ${\bf P}_{ij}$ of the projection operator ${\mathsf
P}$ that couple to the relevant displacement differences. Note that the ${\bf
P}_{ij}$ are constant parameters that only depend on the position
vectors of the reference configuration, ${\bf R}_i$. The size of the
coarse-graining volume $\Omega$ surrounding particle $i$ is set by the
range of the interaction. In the rest of this paper, we take $\Omega$
as the nearest neighbour shell. A positive value of the non-affine field
$h_X$ enhances non-affine distortions of the lattice~\cite{sas1,sas2},
namely, fluctuations in particle positions projected onto a subspace
spanned by those eigen-distortions of $\Omega$ that cannot be represented
as combination of affine deformations.

We describe below our network model. The reference lattice structure
$\{{\bf R}_i\}_{i=1, ..., N}$ is an ideal triangular lattice. The
Hamiltonian of this model is that of a standard network of point non
self-avoiding vertices connected by harmonic bonds~\cite{network},
\begin{eqnarray} 
{\cal H}_0 &=& \sum^N_{i=1} \frac{{\bf p}_i^2}{2 m} + 
\frac{K}{2} \sum_{i=1}^N\sum_{j\in\Omega,i<j} 
(|{\bf r}_j - {\bf r}_i| - |{\bf R}_j - {\bf R}_i|)^2 
\nonumber \label{hamil}
\end{eqnarray}
with ${\bf p}_i$ the momentum, $m$ the mass, ${\bf r}_i$ the instantaneous
position, and ${\bf R}_i$ the reference position of vertex $i$ as before.
The length scale is set by the lattice parameter $l$, the energy scale
by $K l^2$, and the time scale by $\sqrt{m/K}$. We use those as our units in the following, effectively setting $l =
m = K = 1$. A dimensionless inverse temperature is given by $\beta =
K l^2/k_B T$, with $k_B$ the Boltzmann constant.

For some of the calculations reported here, we attach finite sized repulsive particles with every vertex. The Hamiltonian is therefore augmented
to ${\cal H}_0' = {\cal H}_0 + {\cal H}_{WCA}$ with
\begin{eqnarray}
{\cal H}_{WCA} &=& \sum_{i=1}^{N-1} \sum_{j>i} v_{\rm WCA}(r_{ij})
\label{hamil_LJ}
\end{eqnarray}
The interaction potential for a pair of particles, separated by a distance
$r$, is
\begin{equation}
v_{\rm WCA} = 4\phi\left[
  \left(\frac{r_{0}}{r}\right)^{12} 
- \left(\frac{r_{0}}{r_{c}}\right)^{12}
- \left(\frac{r_{0}}{r}\right)^{6} 
+ \left(\frac{r_{0}}{r_{c}}\right)^{6}
\right]
\end{equation}
for $r \leqslant r_{c}=2^{\frac{1}{6}} r_{0}$ and $v_{\rm WCA}=0$ for
$r>r_c$. We use $\phi=1$ and $r_{0}=0.6 l$, respectively.

\subsection{Simulation Details}
\label{Mod_N_SIM_det}
We perform molecular dynamics simulations as well as Monte Carlo
in combination with sequential umbrella sampling of the regions of
configuration space that are otherwise inaccessible using simple MC or
MD simulation techniques.

\paragraph*{\bf Molecular Dynamics.} MD simulations in the canonical
ensemble, i.e.~at constant number $N$ of vertices, area $A$ and
temperature $T$, were done using a leapfrog algorithm, coupling the
system to a Brown and Clarke thermostat~\cite{allen,frenkel}. The size
of the systems ranges from $N=100$ to $N=40000$ vertices. Typically,
unless otherwise stated, we used an MD time step of $\delta t = 0.002$
and inverse temperature $\beta=200$. These parameters are the same even when repulsive, WCA particles are attached to the vertices. Typically, the solid was held for $5\times10^5$ MD steps followed by the collection of data for further $5\times10^5$ MD steps at an interval of $1000$ steps.

\paragraph*{\bf Sequential Umbrella Sampling.} Standard Metropolis Monte
Carlo~\cite{frenkel,binder} is inefficient in sampling systems with
free energy barriers separating different regions in configuration
space. SUS is an advanced sampling technique~\cite{SUS,frenkel,binder}
that ensures good sampling of the entire range of pertinent states. Our
implementation of SUS-MC in the $NAT$ ensemble involves dividing the range
of the relevant order parameter, $X$, into small windows to be sampled
successively starting at $X=0$. Histograms denoted by $H(n)$ keep track
of how often each value of $X$ within the $n$th window is realized, with
$H(n)_{L}$ and $H(n)_{R}$ representing the left and right boundaries of
the $n$th histogram. Now, a predetermined number of MC moves are attempted
per window. MC moves resulting in $X$ within the chosen window are accepted
or rejected using the conventional Metropolis criterion and the relevant
histograms are modified accordingly. Any moves leading to values of $X$
outside the chosen window are rejected with appropriate modification of
the histograms at the boundaries to ensure detailed balance~\cite{SUS}.
Finally, the un-normalized relative probability distribution of $X$
can be computed using,
\begin{equation*}
\frac{P(X_{n})}{P(X_{0})}=
\frac{H(0)_{R}}{H(0)_{L}} \cdot \frac{H(1)_{R}}{H(1)_{L}} \cdots
\frac{H(n)_{R}}{H(n)_{L}}.
\end{equation*}
The SUS-MC runs, for the network model, were done for systems with $N =
900$ at $\beta=200$ and density $\rho=1.1547$ (which corresponds to our choice $l=1$ for the lattice parameter).
For systems with $N=900$, we considered the range between $X=0.0$
and $X=1.0$ and divided this into $500$ sampling windows with $8\times 10^{7}$
MC moves attempted in each window. In each MC move, maximal particle displacements
of $0.2\,l$ along the $x$ and $y$ directions are allowed. Apart from
simulations with $N=900$ vertices, systems with $N=100$ and $N=400$
were studied using SUS-MC. As the variance of the parameter $X$
is proportional to inverse of the system size~\cite{sas2}, the range of
$X$ for sampling was chosen accordingly, viz.~$0.0 \le X \le 3.0$
for $N=100$, $0.0 \le X \le 1.5$ for $N=400$, and $0.0 \le X \le 1.0$ for
$N=900$.  Also for $N=100$ and $N=400$, the range of $X$ was divided into
500 sampling windows with $8\times 10^{7}$ MC trial moves in each window.

\section{Results}
\label{sec2}
\subsection{Crystal properties in the presence of small $h_X$}
We first study the properties of our model solid in the low temperature limit, 
where a harmonic approximation becomes exact
{\em even in the presence of the non-affine field} $h_X$. Indeed, the
complete low-$T$ statistical mechanics of the system can be obtained as
long as the periodic crystalline phase is stable.  Therefore we
begin by first presenting analytic results for finite values of
$h_X$~\cite{sas1,sas2,sas3} in the ideal crystal. These calculations
provide an estimate of the critical value, $h_X^0$, at which the crystal
becomes unstable under the application of the non-affine field.

\begin{figure}
\begin{center}
\includegraphics[width=0.4\textwidth]{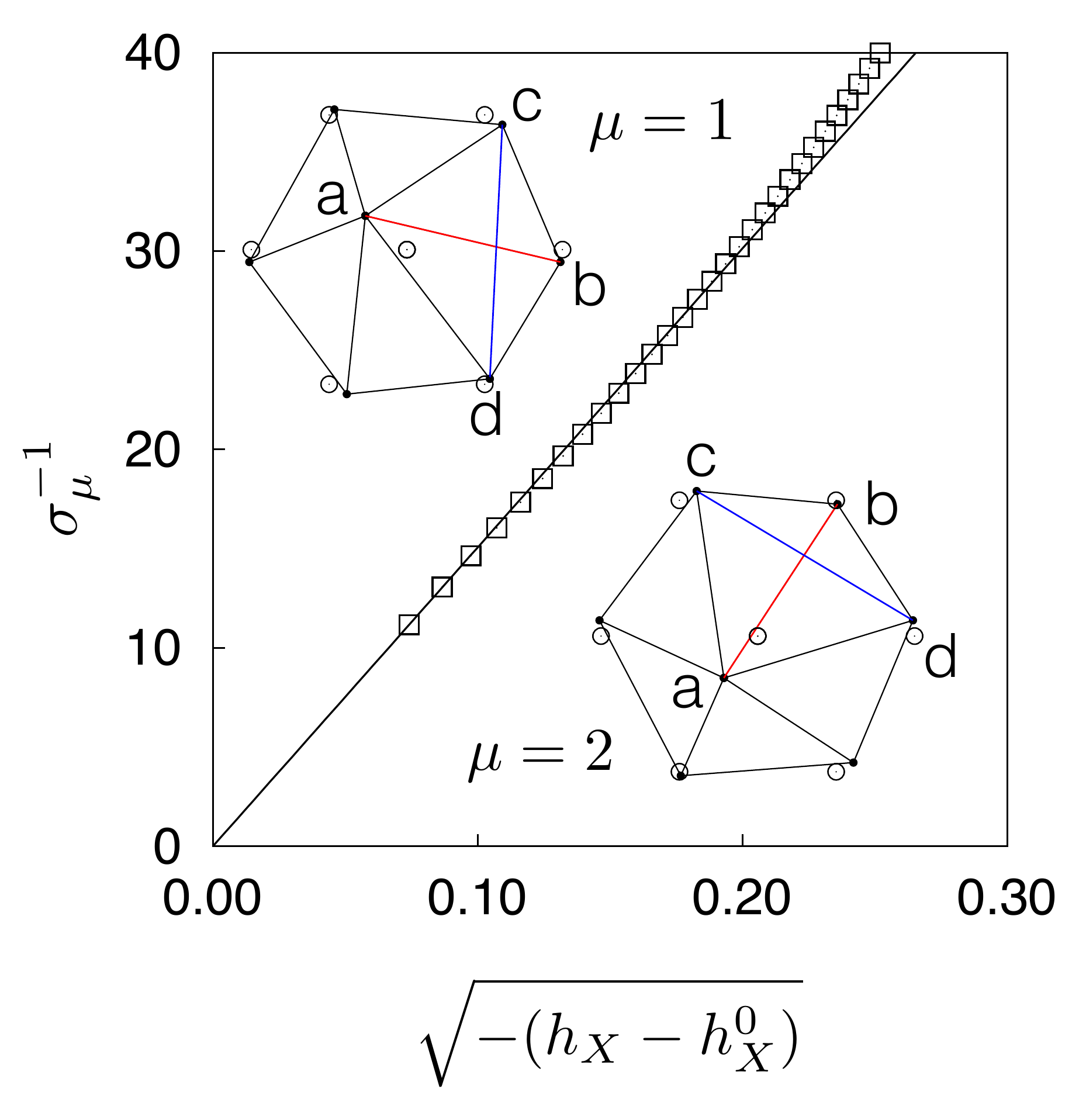}
\caption{\label{fig1} Plots of inverse eigenvalues $\sigma_1^{-1}$
and $\sigma_2^{-1}$ as a function of $h_X$, corresponding to the two
degenerate non-affine modes at the lowest energy (filled squares).
{\color{black}  The solid curve is a linear fit through the last three
data points. The lattice distortions corresponding to these modes are
illustrated in the inserted sketches. Note that in each case nearest
neighbours $a,b$ (red lines) move apart while next nearest neighbours,
$c,d$ (blue lines) come closer (see text).}}
\end{center}
\end{figure}

In order to understand how crystal properties are affected by the
non-affine field, $h_X$, we calculate the eigenvalues of the dynamical
matrix~\cite{born,harmdyn} and the corresponding eigenvectors. The
dynamical matrix is obtained from the Fourier transform of the Hessian
${\cal D}({\bf R},{\bf R}') = \partial^2 {\cal H}/\partial {\bf u}({\bf
R})\partial {\bf u}({\bf R}')$. The full expression for ${\cal D}$
in the presence of a non-affine field $h_X$  can be worked out in a
straightforward manner and one obtains,
\begin{equation*} 
\mathcal{D}= K \left( \begin{array}{cc}
3 -2 \mathcal{A}_{1}- \mathcal{A}_{2} - 
\frac{2 h_X}{K}\mathcal{A}_{X}&\sqrt{3} \mathcal{A}_{3} \\
\sqrt{3} \mathcal{A}_{3} & 3 -3 \mathcal{A}_{2} - 
\frac{2 h_X}{K}\mathcal{A}_{X}\end{array}\right),
\end{equation*}
with
\begin{eqnarray*}
\mathcal{A}_1 & = & \cos(q_{x}), \\ \nonumber
\mathcal{A}_2 & = & \cos(\frac{1}{2}q_{x})\cos(\frac{\surd3}{2}q_{y}), \\ \nonumber
\mathcal{A}_3 & = & \sqrt{3}\sin(\frac{1}{2}q_{x})\sin(\frac{\surd3}{2}q_{y}), \\ \nonumber
\mathcal{A}_{X}&=&\frac{1}{6}\Bigg\{ 
\left[60-28\cos(q_{x}l)-56\cos\Big(\frac{q_{x}l}{2}\Big)
\cos\Big(\frac{q_{y}\sqrt{3}l}{2}\Big) \right] \nonumber\\
&& + \left[4\cos(2q_{x}l)+8\cos(q_{x}l)\cos(q_{y}\sqrt{3}l)\right] 
\nonumber \\
&& + \left[8\cos\Big(\frac{q_{x}3l}{2}\Big)
\cos\Big(\frac{q_{y}\sqrt{3}l}{2}\Big)+4\cos(q_{y}\sqrt{3}l)\right]
\Bigg\}. \nonumber 
\end{eqnarray*}
Note that the leading order term in $\mathcal{A}_{X}(q)$ is of order
${\mathcal O}(q^4)$, so that $h_X$ does not contribute to the speed of
sound or to elastic constants.

Increasing $h_{X}$ enhances $\langle \chi \rangle$ locally,
which is associated with a softening of certain phonon modes. At small
${\bf q}$, the transverse phonon modes are softened showing that
the solid becomes nearly unstable to large wavelength shear modes.
We find that most of this contribution comes from the two softest non-affine modes, i.e.~the eigenvectors of ${\mathsf P}{\mathsf C }{\mathsf P}$ with the smallest eigenvalues $\sigma_\mu$ ($\mu = 1, 2$).  These two eigenvalues are identical and the corresponding modes are illustrated in the
sketches of Fig.~\ref{fig1}: distortions with respect to the reference
configuration (open circles) are generated where nearest neighbours $a,
b$ move away from each other while next-nearest neighbours $c,d$ come
closer tending to nucleate a dislocation dipole~\cite{sas2}.  Note that
any linear combinations of these modes are also degenerate and so the
direction of the modulation wavenumber, with magnitude $\chi^{-1/2}$,
varies in space pointing randomly in all directions consistent with
crystal symmetry.

In Fig.~\ref{fig1}, we plot the reciprocal of the largest non-affine
eigenvalue ($\sigma_{\mu}$ with $\mu = 1, 2$) as a function of
$h_X$. As $h_X$ increases, this eigenvalue vanishes as $\sqrt{h_X -
h_X^0}$ pointing to an underlying saddle-node bifurcation point beyond
which a crystalline solid cannot exist~\cite{sas1}. Similar behavior
is shown by all the eigenvalues (not shown) corresponding to the
non-affine modes~\cite{sas2}. As we can estimate from the plot, the
eigenvalues $\sigma_{\mu}$ with $\mu = 1, 2$ diverge around $h_X^0 =
0.072$ {\color{black} -- the limit of stability of the crystal in a
non-affine field}.

\subsection{Pleated configurations at $T = 0$}
\begin{figure}
\begin{center}
\includegraphics[width=0.48\textwidth]{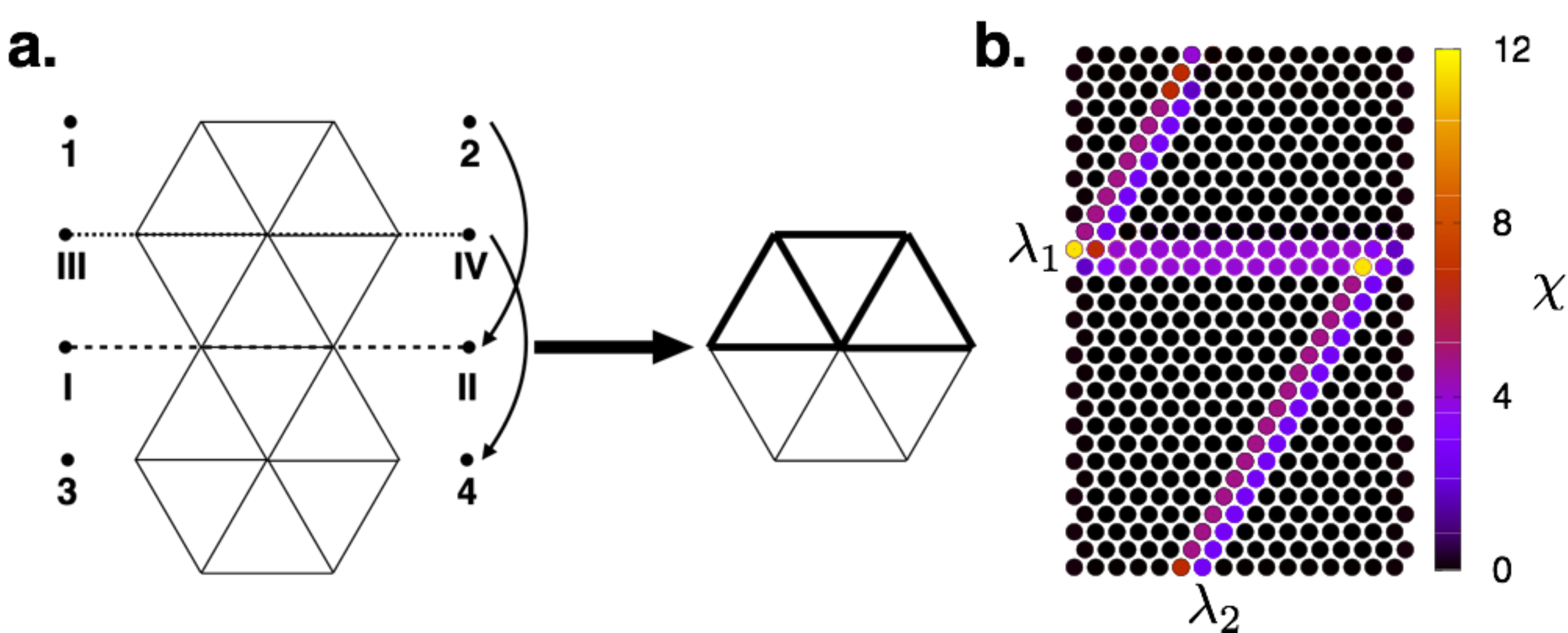}
\caption{\label{pleat} {\bf a.} Schematic illustration of a pleat
in the triangular lattice as an origami fold. A ``valley''
fold at $\mathsf{I}-\mathsf{II}$ and a ``mountain'' fold at
$\mathsf{III}-\mathsf{IV}$ bring the vertices at $\mathsf{I},\mathsf{II}$
and $1,2$ as well as $\mathsf{III}, \mathsf{IV}$ and $3,4$ together to
obtain the final two-dimensional configuration shown on the right. Note
that the pleated  configuration has two layers less and $7$ overlapped
bonds together with $5$ overlapped vertices. Such an operation can
be continued along a crystalline row of vertices creating a complete
pleat. Note further that none of the bonds in the pleat are either
stretched or compressed. {\bf b.} An originally $18 \times 32$ lattice
with one horizontal and one $60^\circ$ pleat of amplitudes $\lambda_1 =
\lambda_2 = 2$ lattice spacings. The vertices are shown as filled
circles coloured according to the value of local $\chi$. The pleats do
{\it not} destroy local crystalline order but have large local $\chi$
values (color bar).}
\end{center}
\end{figure}

Having considered the ideal crystalline ground state of the
non-self-avoiding triangular network, we now ask whether additional
low-energy configurations exist. Indeed, we show that configurations
containing one or more {\em pleats}, as illustrated in Fig.~\ref{pleat},
are also possible. Within a pleat, two rows of vertices overlap
producing a band of twice the local stiffness. Note that the pleated
configuration remains two-dimensional. While similar pleated structures
have been reported~\cite{network,pleats1,pleats2} for such networks
under compression or with disorder, to the best of our knowledge, the
existence of these states for regular crystalline networks has never been
commented upon before. As explained in Fig.~\ref{pleat}a., in a pleated
state no bond is stretched or compressed. A pleated row of vertices
does not destroy local crystalline order and can be distinguished only
by a high value of local $\chi$, pointing out that a finite non-affine
displacement is necessary to produce a pleat. The displacement ${\bf u}$
however becomes a {\em multi-valued} function of the coordinates at the
location of the pleats with each of the two values corresponding to the
two distinct ``leaves'' of the pleat. Since the lattice is stiffer at
the pleats these regions are also regions of enriched stress.

A positive $h_X$ encourages the formation of pleats. In a solid of size
$L$, internal strains $\varepsilon_{int} \sim 1/L$, however, need to
be introduced to fit the the pleated lattice back into the box, making
configurations with a large number of pleats energetically unfavourable.
Consider, for the moment, only pleats of full rows, all in the same
direction, then having $p$ pleats requires a strain of $p/L$ elsewhere,
hence elastic energy $\sim L^2 (p/L)^2 ~\sim p^2$. However, we gain
an energy $\sim h_X p L$ from the field term as $\chi$ is increased to
${\mathcal O}(1)$ in an area of order $p L$. Equating the two predicts
$p \sim h_X L$ or $p/L \sim h_X$, i.e.~a finite fraction of pleats that
increases linearly with $h_X$. At $h_X = 0$ one has competition between
energy $\sim p^2$ and entropy $\sim p \ln L$ and thus expects $p \sim
\ln L$, hence $p/L \to 0$ as $ L \to \infty$.

In Fig.~\ref{pleat}b, we illustrate this with two pleats, one horizontal
and the other tilted at an angle of $60^{\circ}$.  We create these
configurations by shifting rows of vertices either downwards or to
the left by amounts $\lambda_1$ and $\lambda_2$ respectively. Here
$\lambda_1 = \lambda_2 = 2$ lattice spacings. The local $\chi$ is a
quadratic function of both $\lambda_1$ and $\lambda_2$. Note that internal
strains and the term proportional to $h_X$ in ${\mathcal H}$ determine the
relative stability of the pleated configurations. The network responds to
$h_X$ by either increasing the density of the pleats or by introducing
side branches at $60^{\circ}$ (or equivalently $120^\circ$) to the
horizontal. Large $h_X$ favours a large number of pleats and in that
regime several configurations may have the same average non-affineness
$X$ and, at the same time, be degenerate in energy. In the next section
we show that these pleated configurations of the crystalline network
survive at non-zero temperatures and lead to interesting phase behaviour.

\subsection{The phase transition at $T > 0$}
In this section we study in detail the phase transition from an un-pleated
solid to one with pleats at finite temperatures.  This transition can be
located by SUS-MC which gives at a given value of $h_X$ the probability
$P(X)$ to find the system in a state with a certain value of $X$.
The logarithm of this probability is directly related to the corresponding
free energy, $F(X) = - k_B T \ln P(X) + C$ (with $C$ a constant).

\begin{figure}[h!]
\begin{center}
\includegraphics[width=0.5\textwidth]{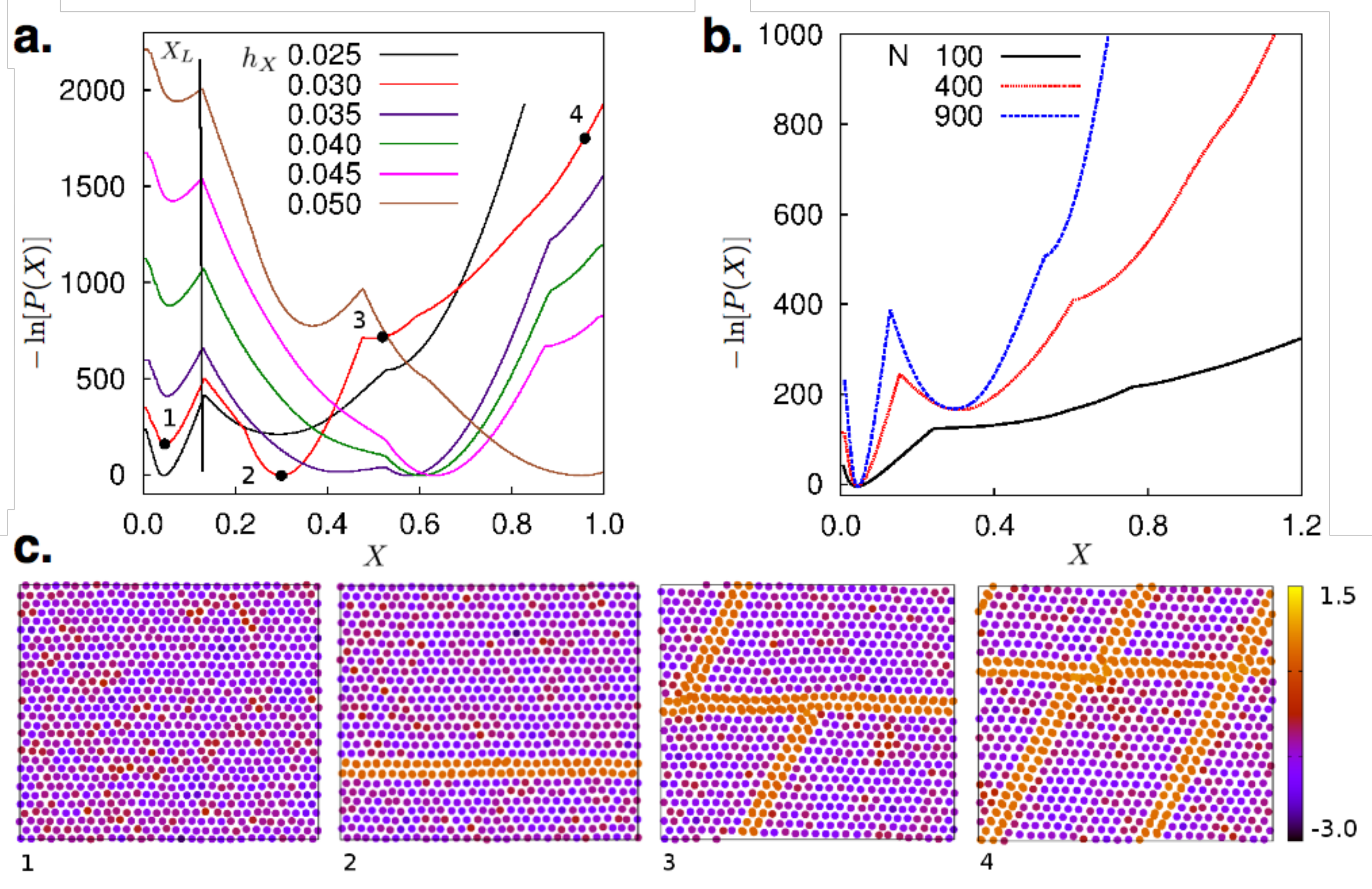}
\caption{\label{fig3} {\bf a.} $-\ln(P(X))$ obtained from sequential
umbrella sampling of a $30\times30$ lattice at various values of
$h_X$. The first minimum always corresponds to the ideal triangular
lattice. As $h_X$ increases, the crystalline solid is destabilised with
respect to other minima at larger $X$. The vertical black line marks
the position of the first saddle point showing that it is approximately
independent of $h_X$. {\bf b.} $-\ln(P(X))$ obtained from sequential
umbrella sampling of $10\times10$, $20\times20$, $30\times30$ and
$50\times50$ lattices at $h_{X}=0.025$ showing that the $P(X)$ becomes
sharper with increasing system size. {\bf c.} Particle configurations at
specific points of $-\ln(P(X))$ for the $30\times30$ solid as indicated
by dots in {\bf a}. The colours represent the local values of $\log_{10}
\chi$.}
\end{center}
\end{figure}

Figure~\ref{fig3}{\bf a} displays $-\ln P(X)$ for the system
with $N= 30\times 30$ lattice sites at different values of $h_X$.
All the distributions exhibit a first minimum around $X = 0.05$
consistent with our results for the ideal network in the harmonic
approximation~\cite{sas2}. A saddle point also appears at $X_{\rm
saddle}\approx 0.1$ which is almost independent of $h_X$. For $X>X_{\rm
saddle}$, the function $-\ln P(X)$ first decreases until a second minimum
is reached. At higher values of $X$ more minima and shoulders can be
discerned in the distributions; at high values of $h_X$ these lie below the
first two minima and thus correspond to the stable states in this regime.

In Fig.~\ref{fig3}{\bf b}, $P(X)$ at $h_X = 0.025$ for various
system sizes $N$ is plotted. In a conventional first-order phase
transition~\cite{CL}, $P(X)$ sharpens with increasing system size such
that $-N^{-1}\ln(P(X,N))$ approaches its thermodynamic limit. In our
system, while $P(X)$ does become narrower with increasing $N$, detailed
features of $P(X)$ depend non-trivially on the system size. Indeed
for larger system sizes, additional minima appear that correspond to domain
configurations not possible for smaller sizes. One of the great
advantages of the SUS-MC method is that configurations that contribute
to $P(X)$ in each range are directly available.

Snapshots for different values of $X$ at $h_X = 0.030$ (as marked in
Fig.~\ref{fig3}{\bf a} by black dots) are displayed in Fig.~\ref{fig3}{\bf
c}. Here each  vertex is represented as a filled circle and coloured
according to the local value of $\log_{10}\chi$. While the configuration
corresponding to the first minimum in $-\ln(P(X))$ is a homogeneous crystal,
the one corresponding to the second minimum is an inhomogeneous phase
where a band of vertices with a high positive $\chi$ value percolates
through the crystal. The latter band does not simply consist of two
lattice rows, as the snapshot may suggest.  While the two rows fit
perfectly into the hexagonal structure, each of the two rows consist
of a pair of overlapping rows, i.e.~a pleat. At higher values of $X$,
in addition to the horizontal pleat, side branches form at an angle of
$60^\circ$ with a geometry similar to that used in the $T=0$ calculation.
On further increasing $X$, the side branches cross the main band such
that they also percolate through the system.

\begin{figure}
\begin{center}
\includegraphics[width=0.45\textwidth]{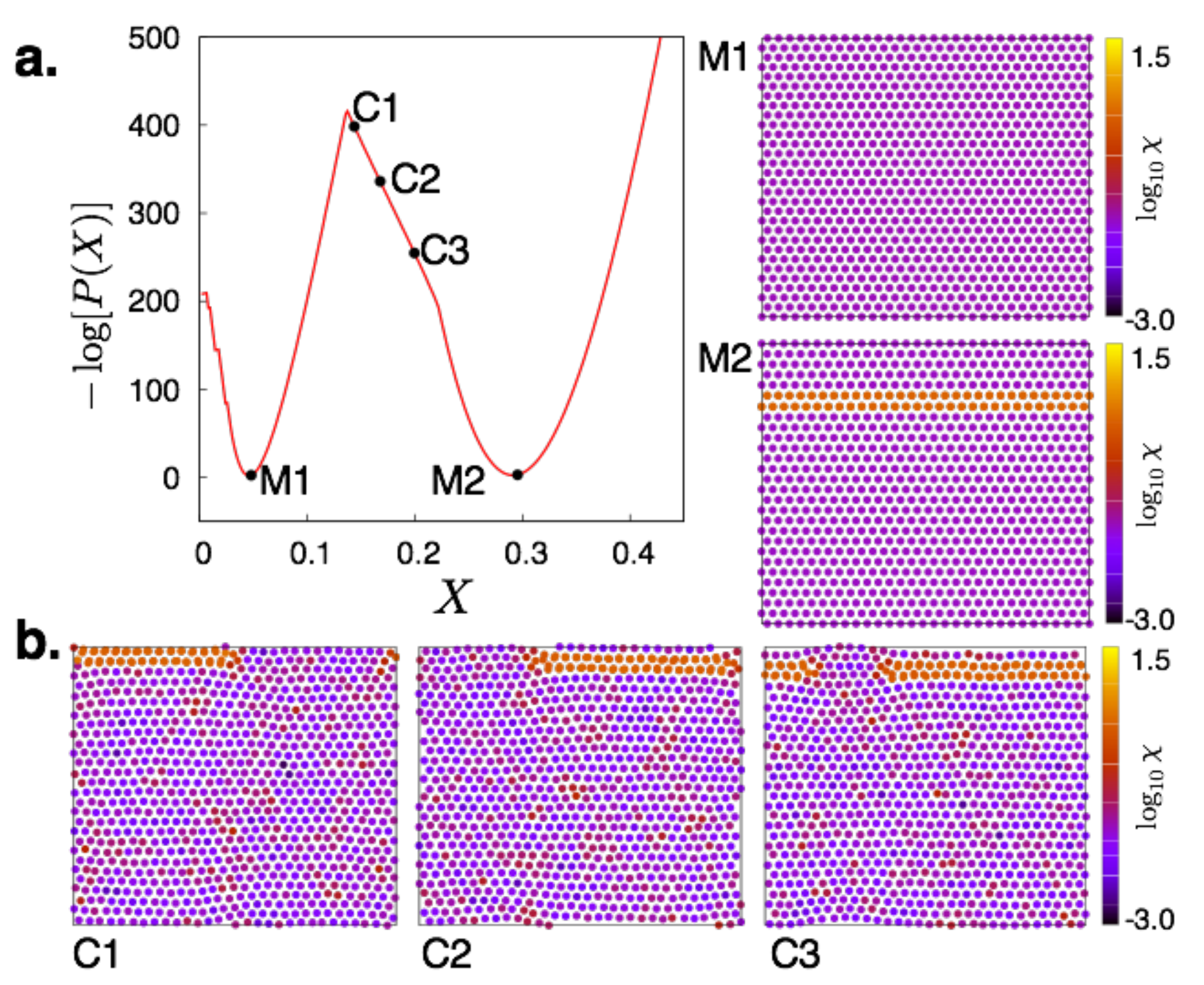}
\caption{\label{fig4} {\bf a.} $-\ln P(X)$ is plotted as a function of
$X$ at $h_X = 0.027$, where the crystal and the inhomogeneous phase
with a non-affine stripe of particles are at coexistence. {\bf b.}
Configuration snapshots with $\log_{10} \chi$ colormap corresponding to
the states indicated in  {\bf a.} While the snapshots C1, C2,
C3 represent instantaneous configurations, the snapshots M1 and M2 were
obtained from a time average over $10^{4}$ subsequent configurations.}
\end{center}
\end{figure}

Is there a finite temperature phase transition from a homogeneous
crystalline phase to an inhomogeneous phase with a pleat of ``non-affine''
vertices? At such a transition, the probability distribution
$P(X)$ would exhibit two peaks with the area under both peaks being equal
\cite{binder}. As one can infer from Fig.~\ref{fig3}{\bf a}, this happens
for a value of $h_X$ between 0.025 and 0.030. To obtain an estimate of
the coexistence value for $h_X$, one can use histogram reweighting and
deduce from a reference distribution at $h_X^{(1)}$, $P(X, h_X^{(1)})$,
the distribution at $h_X^{(2)}$ via
\begin{equation}
- \ln P(X, h_{X}^{(2)}) = - \ln P(X, h_{X}^{(1)}) 
+ \beta \left[ h_{X}^{(2)} - h_X^{(1)} \right]  NX \, .
\label{eq_reweight}
\end{equation}
Eq.~(\ref{eq_reweight}) may now be used to determine the distribution
$P(X, h_X^{\rm coex})$ for which the area under the peaks corresponding to
the coexisting phases is equal. We accomplish this using an iterative
procedure where successive refinements of $h_X^{\rm coex}$ are
obtained from SUS-MC simulations at a previously estimated values of
$h_X^{\rm coex}$.  From this procedure we find $h_X^{\rm coex} = 0.027$.
We check that the final  $P(X, h_X^{\rm coex})$ shows two peaks with equal
area under them as required for co-existence (see Fig. \ref{fig4}{\bf
a}). States in the two-phase region (e.g.~those points marked as C1,
C2, and C3 in the figure) are also visible. Here, the two phases are at
coexistence and separated from each other by an interfacial region. As
the snapshots for the states C1, C2, and C3 show the pleat does not
percolate through the system in the two-phase region and remains as
a droplet terminated at two opposite vertices by the presence of the
homogeneous crystal phase. Local $\chi$ is a convenient collective
variable useful for characterising pleats. However, the thermodynamic
variable which ought to show the interface between the homogeneous
crystal and the inhomogeneous pleated phase is the space dependent,
complex amplitude of the appropriate vertex density modulation. This
is computationally difficult to obtain. Fortunately, as we show below,
the local (uniaxial) stress distribution works just as well.

\begin{figure}
\begin{center}
\includegraphics[width=0.47\textwidth]{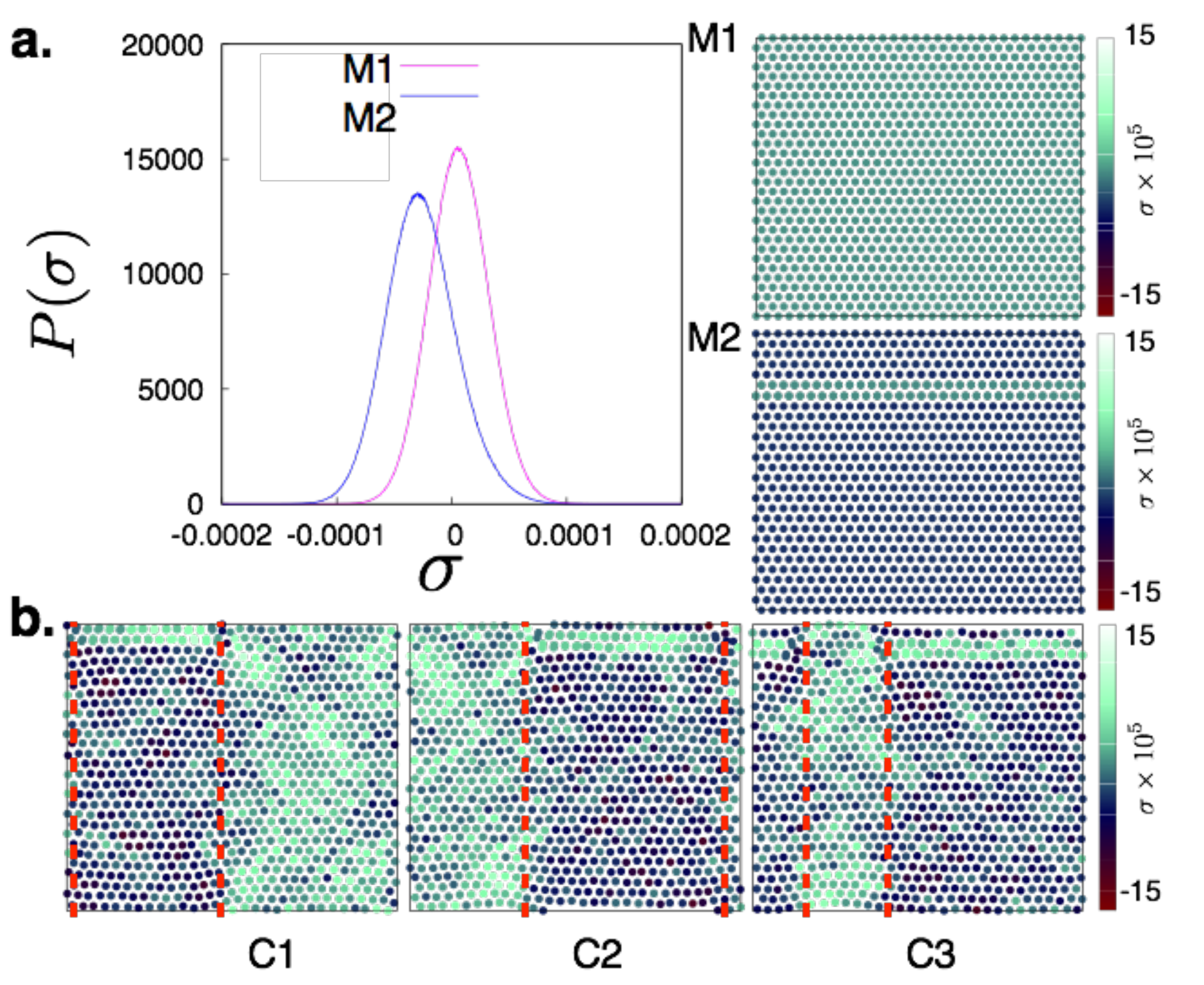}
\caption{\label{fig5} Distribution of local stresses $P(\sigma)$
corresponding to the minima M1 and M2, indicated in Fig.~\ref{fig4}. The
snapshots correspond to those in Fig.~\ref{fig4} but now the colormap
represents the local stresses.}
\end{center}
\end{figure}

In Fig.~\ref{fig5} we demonstrate that the two phases at coexistence (Fig.~\ref{fig4}) differ from each
other with respect to the distribution of local stresses~\cite{frenkel} at coexistence,
$P(\sigma)$, where $\sigma = \sigma_{xx} - \sigma_{yy}$, the {\em deviatoric} or {\em uniaxial} stress component. 
The distribution for the inhomogeneous phase (M2) has
its maximum at the negative value $\sigma = - 4 \times 10^{-5}$ and
is asymmetric, {\color{black}with a pronounced excess contribution at
positive values of $\sigma$. The latter asymmetry reflects
the localization of stresses in the pleat with a high value of $\log_{10}
\chi$} (see snapshot corresponding to M2 in Fig.~\ref{fig4}). In contrast,
the distribution corresponding to M1 is symmetric and its maximum
is at the positive value $\sigma = 2 \times 10^{-5}$. The corresponding
snapshot indicates a homogeneous distribution of stresses.  Thus, the
two phases at M1 and M2 differ with respect to the average stress value
and the localization of stress. In the M2 phase, the average stress
value in the pleated region where the stress is localized is similar
to that of the homogeneous M1 phase.  All this is reflected in the
snapshots corresponding to the two-phase region (C1, C2, and C3). Now
the interfaces between the coexisting phases are clearly visible (dashed
red lines in the plot of configuration C1, C2 and C3).  While in the M1
phase stress is distributed throughout the crystal, in the M2 phase,
it is eliminated from the un-pleated part and concentrated mostly at
the pleat. There are two interfaces due to periodic boundary conditions
and the amount of the two phases at a given state is controlled by
the lever rule.  Consequently, the free energy decreases linearly
from state C1 to C3 (see Fig.~\ref{fig4}), although the total area of
the interfaces is constant for these three states. In the periodically
repeated system, the M2 phase becomes a vertical  stripe of undistorted
network punctuated by a parallel array of horizontal pleats at regular
intervals.  Thermal undulations along the interfaces are also visible.
In summary, Fig.~\ref{fig5} demonstrates that the order parameter of the
transition from the homogeneous crystal to one with stress localisation
within pleats can be connected to distribution of local stresses. We end this section with a comment on $P(\sigma)$ from our simulations. Fig.\ref{fig5} suggests that the $\sigma \to -\sigma$ symmetry is broken at $\varepsilon = 0$. To obtain a complete description one would need to sample all degenerate, globally rotated copies of the crystal and also evaluate the full stress tensor $\sigma_{ij}$. This is however not necessary for our purpose here. 

\subsection{Dynamical transition and plastic deformation}
In the last section we studied the properties of the pleated
configurations showing that they form the stable equilibrium phase
beyond a first order transition from an un-pleated to a pleated phase, which occurs in
our network solid at $h_X = h_X^{\rm coex} \approx 0.027$ when $\beta=200$.  How can
such configurations form dynamically? We turn now to study this kinetic
transition.

We have mentioned before that pleated configurations imply a multi-valued
displacement field. Obviously, such a configuration cannot be represented
as a linear combination of hydrodynamic phonon fluctuations of the
network. Pleated configurations therefore need to form by the nucleation
and growth of non-hydrodynamic, localized droplets. Incomplete pleated
regions surrounded by a strained network have been observed and described
in detail in the last subsection (see for example the C1 configuration in
Fig.~\ref{fig4}). For values of $h_X$ at which pleated states become
globally stable, such droplet configurations, which lie in the saddle
region in-between minima corresponding to un-pleated and pleated states
cost extremely high free energy. Such high barriers prevent the equilibrium
transition from occurring in MD simulations.

\begin{figure}
\centering
\includegraphics[width=0.48\textwidth]{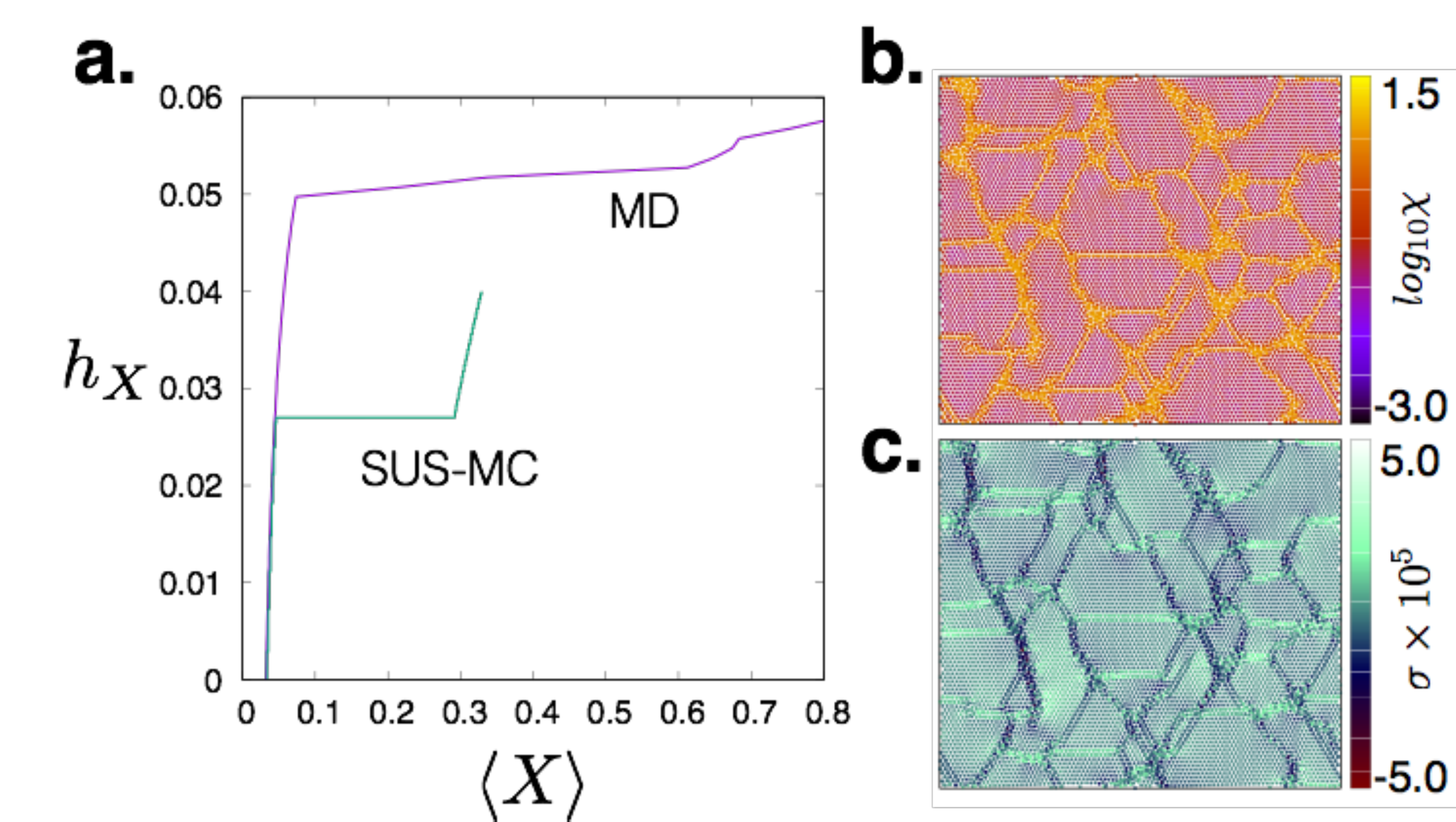}
\caption{{\bf a.} $h_X$ as function of $\langle X \rangle$ for both
SUS-MC and MD simulations. {\bf b.} and {\bf c.} Configurations with
vertices coloured according to $\log_{10}\chi$ and $\sigma\times 10^{5}$
obtained after the dynamical transition at $h_X=0.06$. \label{md-sus}}
\end{figure}

A dynamical transition at $T > 0$ to a pleated configuration is possible
only when $h_X$ becomes sufficiently large so that the lattice is close
to being (but not quite!) locally unstable and the free energy barrier
is substantially reduced. In Fig.~\ref{md-sus}b, we have plotted $h_X$
against $\langle X \rangle$ obtained from SUS-MC. The $\langle X \rangle$
values were obtained by a histogram reweighting method. Together with
these results, we have also plotted results from MD simulations of the
same network where $\langle X \rangle$ now represents an average over
the MD simulation time. The MD and the SUS-MC results both show a jump in
$\langle X \rangle$ at the pleating transition. However, the transition
in MD occurs at a much larger value of $h_X$ showing that for a large
range of $h_X$, the un-pleated state remains metastable. Note also that
the value of $\langle X \rangle$ in the un-pleated state just before the
dynamical transition is roughly equal to the value at the saddle point
as shown in Fig.~\ref{fig3}. A Lindemann like criterion~\cite{linde,
CL} viz.~$X = X_L = X_{\rm saddle}$ just below the transition in the
crystal phase is thus operative at this kinetic transition. Finite size
effects in the MD simulations roughly follow those in the MC consistent
with the shift of the position of the saddle point to smaller $X$ values
(see Fig.~\ref{fig3}{\bf b}) with increasing $N$.

Configurations obtained just after the transition, at $h_X=0.06$, are plotted in
Fig.~\ref{md-sus}{\bf b} as both local $\chi$ and $\sigma$ maps.
While pleated regions of higher local stress similar to
the SUS-MC results are also seen here, the arrangement of the pleats is
disordered. Close examination of the configurations also suggests that
some of the pleated regions are amorphous. This may be understood as
follows. As soon as the thermal energy required to cross the free
energy barrier is available, the solid begins to form local pleats. Since
many equivalent pleated states are equally stable at these high values
of $h_X$ deep within the equilibrium phase boundary, the solid locally
chooses between the several degenerate pleated states and relaxes,
typically, to the nearest metastable free energy minimum. Further relaxation to
the true equilibrium ground state, however, now needs large scale
rearrangements of the network. As a
consequence, the pleated solid shows ageing dynamics.

To show this, we compute the overlap function $Q(t) = N^{-1}\sum_i^N
w(|{\bf r}_i(t) - {\bf r}_i(t_w)|)$ for all vertices with local $\chi \geq
\chi_{cut} =1$. The weight function $w(x)$ is zero or $1$ depending on
whether $x > a$ or $ x \le a$, with $a$ being some predetermined length,
smaller than the lattice spacing $l$; we choose $a=0.1l$. For a fixed
value of $h_X = 0.06$, starting from an initial crystalline structure,
the system is allowed to relax for a ``waiting'' time $t_w$, before
$Q(t)$ is computed. To obtain good statistics, $Q(t)$ is averaged over
many independent runs.

\begin{figure}
\begin{center}
\includegraphics[width=0.48\textwidth]{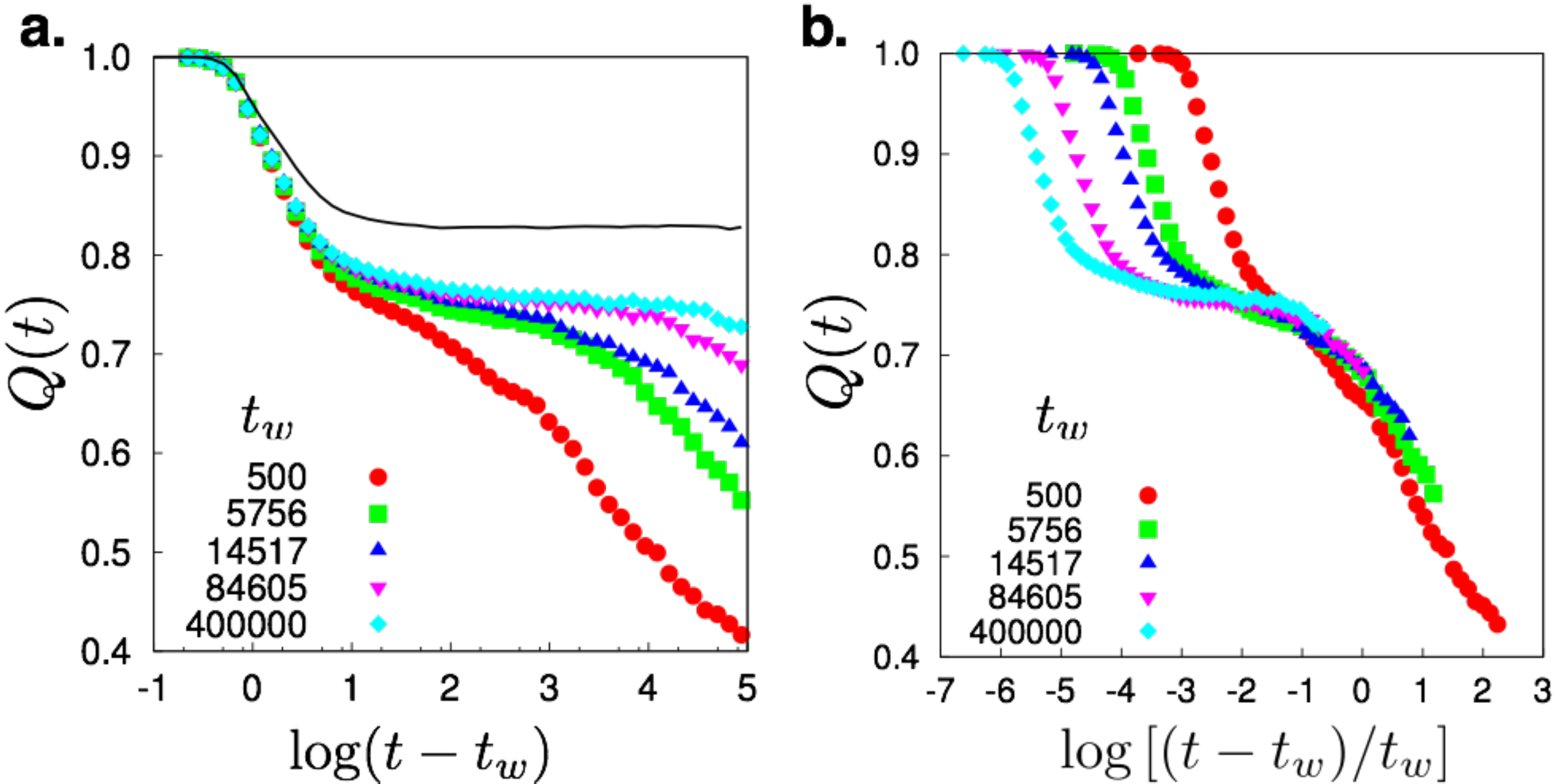}
\caption{\label{fig10} {\bf a.} Ageing plot for the overlap function,
$Q(t)$ as a function of $\ln t$ for several waiting times $t_w$ (see
text). The black curve shows $Q(t)$ for a crystalline lattice i.e. at
a lower value of $h_{X}=0.02$ {\bf b.} Scaling collapse of the overlap
function.  The data is from MD simulations of a $20\times20$ lattice at
$h_{X}=0.06$, averaged over $60$ independent realizations.}
\end{center}
\end{figure}

In Fig.~\ref{fig10}{\bf a}, we plot $Q(t)$ for five values of $t_w$
spanning three orders of magnitude. In each case $Q(t)$ shows an initial
rapid decrease to a plateau value, then slow relaxation in the plateau followed
by an eventual escape away from the plateau at large times. $Q(t)$ depends
on both $t$ and the waiting time $t_w$ implying that the system shows
ageing behaviour. For a process where the system relaxes quickly to a
steady state configuration, $Q(t)$ at long times always has a non-zero
limiting value. To illustrate this, we plot in Fig.~\ref{fig10}{\bf a}
$Q(t)$ for all the vertices in the crystalline network at a lower $h_X =
0.02$. The long time behaviour of $Q(t)$ in this case is very different,
saturating to a constant value. On the other hand, for a system showing
complex dynamics requiring long range (and time consuming) particle
rearrangements  $Q(t)$ relaxes to a plateau first but then eventually
decays to zero, on a timescale that grows with $t_w$. To distinguish
between these behaviours, we plot $Q(t)$ against the scaled time $t^*=(t -
t_w)/t_w$ in Fig.~\ref{fig10}{\bf b}. The data collected over all three
decades of $t_w$ collapse on a single curve which decays to zero at large
values of $t^*$ showing that the phase transition kinetics bears strong
resemblance to relaxation in a complex landscape.

All the results described so far correspond to the pleating transition in
a non-self avoiding, 2d triangular crystalline network. If one thinks of the vertices as colloidal particles, then one might construct such
a system experimentally and apply the non-affine field $h_X$ using dynamic
laser traps in the fashion described in detail in Ref.~\cite{sas2} 
Of course, colloidal
particles are self-avoiding and will have an excluded volume. This
should alter the properties of the pleats. Does it also suppress the
pleating transition completely? We now show that 
while the detailed configuration of the pleats are affected
because particles cannot overlap, this does not change any of the
equilibrium or dynamic results substantially. While a complete overlap
is impossible, particles occupy positions determined by a compromise
between the bonding and the non-bonding, hard-core repulsion.

\begin{figure}
\begin{center}
\includegraphics[width=0.48\textwidth]{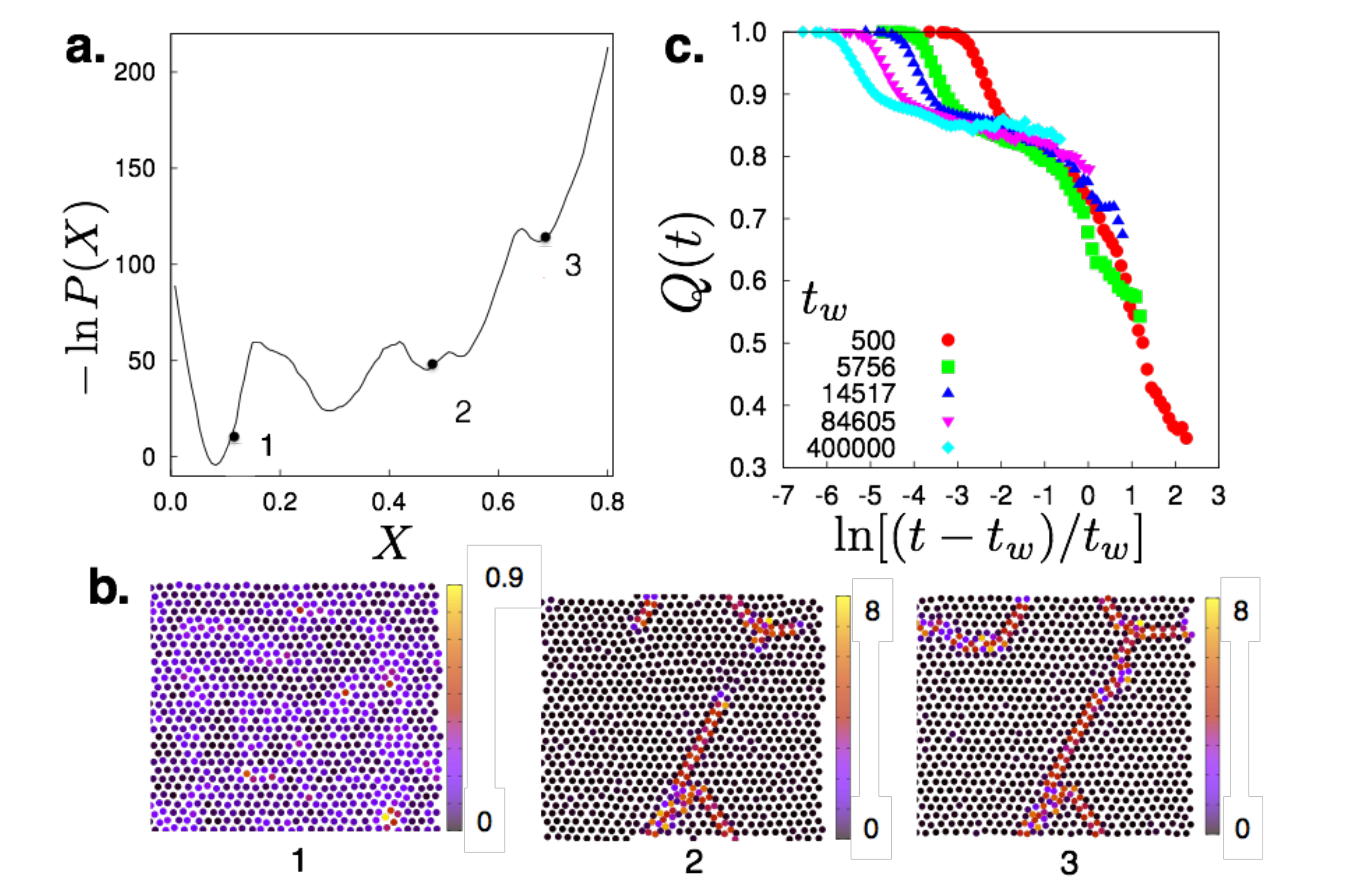}
\caption{\label{fig11} {\bf a.} $-\log P(X)$ at $h_X = 0.054$, obtained
using SUS-MC simulations of a $30\times30$ lattice with vertices occupied
by particles interacting via an additional hard core WCA repulsion. {\bf
b. }Configurations from the SUS-MC simulations at three different values
of $X$ shown in {\bf a.} by black dots. Note that the nature of the pleated
regions is similar to Fig.~\ref{fig3} despite the presence of the hard core repulsion. {\bf c.}
Ageing plot for the collapsed overlap function $Q(t)$ from MD simulations
of a $20\times20$ lattice with the same interactions as in {\bf a.}
The value of $h_X = 0.06$ and the plots were obtained by averaging over
$40$ realisations.}
\end{center}
\end{figure} 

We summarise these results in Fig.\ref{fig11} {\bf a.}-{\bf c}. The
vertices of the network now have, in addition to the harmonic bonded
interaction, a non-bonding interaction, which we modelled using
the purely repulsive WCA potential in (\ref{hamil_LJ}). 

In Fig.~\ref{fig11}{\bf a}, we plot the probability distribution $P(X)$
for a single value of $h_X = 0.054$, close to but lower than the value
at the equilibrium transition. The nature of this curve is very similar
to its counterpart for the non-self avoiding lattice (see Fig.~\ref{fig3}).
However, the pleated states are now somewhat destabilised with respect
to the homogeneous lattice. This is only to be expected because the hard core
repulsion that prevents particle overlaps now makes pleated states with
no bond stretching impossible. This is clear from plots of configurations
in Fig.~\ref{fig11}{\bf b} where particles are seen to come close to one
another without overlaps. In general, then, the position of the equilibrium
transition is shifted to higher values of $h_X$. On the other hand, the
position of the saddle point $X_{\rm saddle}$ is virtually unchanged. As
a consequence, the location of the dynamical transition does not shift
too much so that the equilibrium and dynamical transition points are
now closer to each other.

The qualitative nature of the dynamics is similar and the dynamical
transition for the same system size occurs again for slightly
larger values of $h_X \approx 0.06$. In Fig.~\ref{fig11} we plot the
overlap function $Q(t)$ for this system at a value of $h_X$ after the
transition. Similar ageing dynamics as seen in the network with point
vertices is obtained.

\section{Discussion and conclusions}
In this paper we study a crystalline network  consisting of a lattice of particles (with and without hard core repulsion) permanently
connected to their nearest neighbours by harmonic springs. We show that this undergoes a
phase transition from a homogeneous to a pleated phase {\em provided that
non-affine displacements are artificially enhanced using an external
field}. It is interesting to note that the pleating transition in the
network, where a homogeneous phase with a uniform stress distribution
gives rise to an inhomogeneous phase consisting of an ordered arrangement
of pleats where stress is concentrated, has an analogy in the physics of
Type II superconductors~\cite{tinkham}. Here stress plays the r\^ole of
the magnetic field and the elimination of stress from the un-pleated parts
of the network is a manifestation of a ``stress Meissner effect''. The
possibility of such an effect had been described in the past for crystals
which have irregular modulations~\cite{toledano}. In these crystals, the
order parameter is modulated with a space dependent amplitude and phase
producing a {\em disordered} structure. The non-affine modes discussed
in Section~\ref{sec2}A (see Fig.~\ref{fig1}) do produce a similar
modulation. However, remarkably, the product state is {\em ordered},
contrary to the predictions of Ref.~\cite{toledano}. Strong correlation
effects between the order parameter modulations~\cite{sas2} may be responsible for this
departure. Nevertheless, this  ordered pleated state should be viewed
as a classical version of the Abrikosov vortex lattice. The arrangement
of pleats now performs the same function as the vortices.

There are several ways in which the calculations described here may be
extended. Firstly, we believe that external stress would have significant
effects on the transition observed ere. Specifically, compressive stress,
should decrease the value of $h_X^{\rm coex}$, perhaps even to $h_X^{\rm
coex} = 0$, where the equilibrium transition occurs before the network
becomes locally unstable~\cite{pleats2}. A similar decrease of $h_X^{\rm
coex}$ is possible for uniaxial or shear stresses. Such stresses should
also introduce anisotropy making it possible to design specific pleating
morphologies. Pleating of networks may also have some implications for plastic
deformation in these systems. Preliminary investigations by us do point
to such a possibility and these results will be published elsewhere.

We have confined ourselves, in this paper, to 2d models where
experiments can be performed to check all our predictions with available
technology~\cite{sas2}. As detailed in Ref.~\cite{sas2}, a feedback-loop
may be set up where local particle configurations in a colloidal crystal
may be used to compute non-affine forces which are then administered using
laser traps positioned on-the-fly~\cite{HOT}. We expect such experiments
to yield metastable structures. The enumeration of equilibrium pleated
configurations together with their relative free energies and individual
barriers and transition states as obtained using techniques elaborated
in this paper would, we believe, be useful to analyse the results of
these future experiments.

In principle all our calculations can also be extended in a
straightforward fashion to higher dimensions as pointed out
in ~\cite{sas1}, though experiments on colloids then become more
difficult. Irrespective of this, the analogs of the pleated phase in higher dimensions should certainly be interesting.

\begin{acknowledgments}
We thank S. Ramaswamy, G. Menon, A. K. Sood and C. Dasgupta
for discussions. SS thanks the Okinawa Institute for Science and
Technology for hospitality. SG thanks CSIR India for a Senior Research
Fellowship. Funding from the FP7-PEOPLE-2013-IRSES grant no: 612707,
DIONICOS is acknowledged. PS acknowledges the stimulating research environment provided by the EPSRC Centre for Doctoral Training in Cross-Disciplinary Approaches to Non-Equilibrium Systems (CANES, EP/L015854/1).
\end{acknowledgments}


\begin{thebibliography}{}
\bibitem{origami1}
L. Mahadevan and S. Rica, 
Science {\bf 307}, 1740 (2005).
%
\bibitem{origami2}
D. M. Sussman, Y. Cho, T. Castle, X. Gong, E. Jung, S. Yang, and R. D. Kamien, 
Proc. Natl. Acad. Sci. USA {\bf 112}, 7449 (2015).
%
\bibitem{origami3}
T. Castle,Y. Cho, X. Gong, E. Jung, D. M. Sussman, S. Yang, and R. D. Kamien, 
Phys. Rev. Lett. {\bf 113}, 245502 (2014).
%
\bibitem{irvine}
W. T. M. Irvine, V. Vitelli, and P. M. Chaikin, 
Nature {\bf 468}, 947 (2010).
%
\bibitem{carpets}
N. Geerts and E. Eiser,  
Soft Matter {\bf 6}, 4647 (2010).
%
\bibitem{bayley}
M. A. Holden, D. Needham, and H. Bayley, 
J. Am. Chem. Soc. {\bf 129}, 8650 (2007).
%
\bibitem{shape}
T. Zhang, D. Wan, J. M. Schwarz, and M. J. Bowick, 
Phys. Rev. Lett. {\bf 116}, 108301 (2016). 
%
\bibitem{CL} 
P. Chaikin and T. Lubensky, 
{\it Principles of Condensed Matter Physics} 
(Cambridge Press, Cambridge, 1995).
%
\bibitem{falk}
M. L. Falk and J. S. Langer,  
\pre {\bf 57}, 7192 (1998).
%
\bibitem{sas1}
S. Ganguly, S. Sengupta, P. Sollich, and M. Rao, 
Phys. Rev. E {\bf 87}, 042801 (2013).
%
\bibitem{sas2}
S. Ganguly, S. Sengupta, and P. Sollich, 
Soft Matter {\bf 11}, 4517 (2015).
%
\bibitem{sas3}
A. Mitra, S. Ganguly, S. Sengupta, and P. Sollich, 
JSTAT, P06025 (2015).
%
\bibitem{HOT} 
G. C. Spalding, J. Courtial, and R. D. Leonardo, 
in D. L. Andrews Ed., {\em Structured Light and its Applications} 
(Elsevier, Oxford 2008).
%
\bibitem{network}
D. E. Discher, D. H. Boal, and S. K. Boey, 
Phys. Rev. E {\bf 55}, 4762 (1997).
%
\bibitem{pleats1}
M. F. Thorpe and E. J. Garboczi, 
Phys. Rev. B, {\bf 42}, 8405 (1990).
%
\bibitem{pleats2}
D. H. Boal, U. Seifert, and J. C. Shillcock, Phys. Rev. E, {\bf 48}, 4274 (1993).
%
\bibitem{spectrin}
D. H. Boal, U. Seifert, and A. Zilker,  Phys. Rev. Lett. {\bf 69}, 3405 (1992).
%
\bibitem{spectrin1}
H. Li and G. Lykotrafitis, 
Biophys. J. {\bf 102}, 75 (2012).
%
\bibitem{binder}
K. Binder and D. Heermann,
{\em Monte Carlo Simulation in Statistical Physics:
An Introduction, 5$^{th}$ Ed.}
(Springer, Berlin, 2010).
%
\bibitem{SUS} 
P. Virnau and M. M\"uller, 
J. Chem. Phys. {\bf 120}, 10925 (2004).
%
\bibitem{allen}
M. P. Allen and D. J. Tildesley, 
{\em Computer Simulation of Liquids} 
(Oxford University Press, Oxford, 1987).
%
\bibitem{frenkel}
D. Frenkel and B. Smit,
{\it Understanding Molecular Simulations}
(Academic Press, San Diego, 2002).
%
\bibitem{born}
M. Born and K. Huang,
{\em The dynamical theory of crystal lattices}
(Clarendon Press, Gloucestershire, 1998).
%
\bibitem{harmdyn} 
E. J. Garboczi and M. F. Thorpe,  
Phys. Rev. B {\bf 32}, 4513 (1985).
%
%
\bibitem{linde} 
F. Lindemann, 
Z. Phys. {\bf 11}, 609, (1910)
%
\bibitem{tinkham}
M. Tinkham, 
{\it Introduction to Superconductivity, 2$^{nd}$ Ed.} 
(Dover Publications, New York, 2004). 
%
\bibitem{toledano}
P. Toledano, 
Europhys. Lett. {\bf 78}, 46003 (2007).
%
\end{thebibliography}
\end{document}